\documentclass[twoside]{ilcws08}
\usepackage[latin1]{inputenc}
\usepackage{graphicx}
\usepackage{wrapfig,rotating}
\usepackage{amssymb,amsmath,array}

\begin{document}
\title{
Track Segments in Hadronic Showers: Calibration Possibilities for a Highly Granular HCAL} 
\author{Frank Simon$^{1, 2}$ for the CALICE Collaboration
\vspace{.3cm}\\
1- Max-Planck-Institut f\"ur Physik, F\"ohringer Ring 6, 80805 M\"unchen, Germany
\vspace{.1cm}\\
2- Excellence Cluster 'Universe', Boltzmannstrasse 2, 85748 Garching, Germany
}

\maketitle

\begin{abstract}
The CALICE collaboration has constructed a highly granular hadronic calorimeter based on small scintillator tiles read out with silicon photomultipliers. With this detector, data was taken at CERN and at Fermilab. The high granularity of the detector allows the identification of minimum-ionizing track segments within hadronic showers, demonstrating the imaging capabilities of particle flow calorimeters. These tracks can be used for the cell-by-cell calibration of such a calorimeter. The possibility to calibrate a complete ILC calorimeter with such track segments in the absence of muons is also investigated.
\end{abstract}

\nocite{url}

\section{Introduction: The CALICE Calorimeters}

\begin{figure}
\centering
\includegraphics[width=0.99\columnwidth]{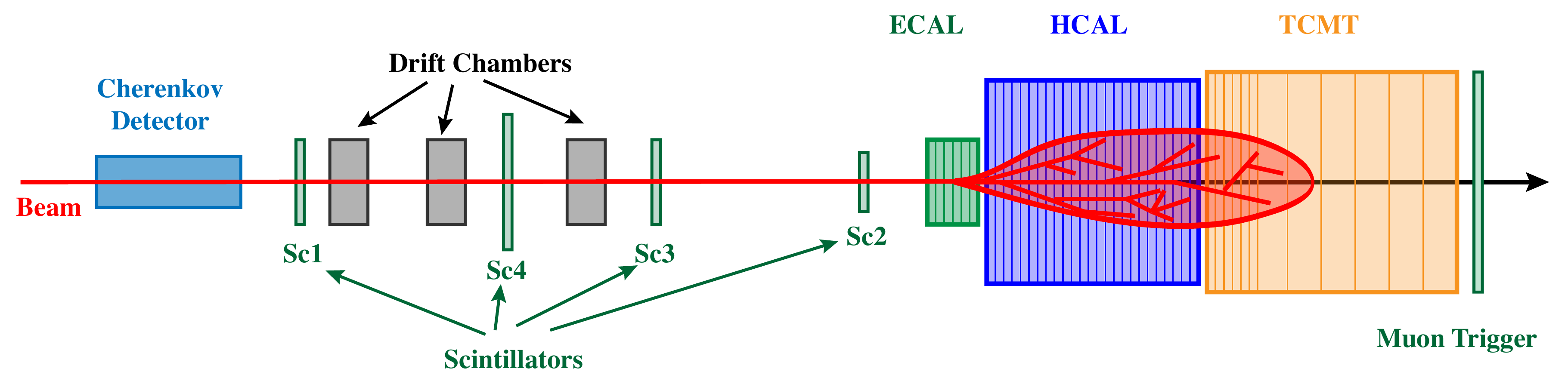}
\caption{Schematic of the CALICE experimental setup at CERN, with electromagnetic and hadronic calorimetry as well as a tail catcher and muon tracker downstream of the calorimeters.}
\label{fig:CALICESetup}
\end{figure}

The goal of the CALICE experimental program is to establish novel technologies for calorimetry in future collider
experiments and to record electromagnetic and hadronic shower data with unprecedented three dimensional spatial resolution for the
validation of simulation codes and for the test and development of reconstruction algorithms. Such highly granular calorimeters are necessary to achieve an unprecedented jet energy resolution at the International Linear Collider \cite{:2007sg} using particle flow algorithms \cite{Thomson:2008zz}.

The CALICE test beam setup \cite{Eigen:2006eq} consists of three separate sampling calorimeters: a silicon-tungsten electromagnetic calorimeter (ECAL), an analog scintillator-steel hadron calorimeter (AHCAL) and a tail catcher/muon tracker (TCMT). The scintillator cells in the latter two detectors are individually read out by silicon photomultipliers (SiPMs) \cite{Bondarenko:2000in}. This setup has been tested extensively in electron, muon and hadron beams at CERN and at Fermilab. Figure \ref{fig:CALICESetup}  shows the schematic setup of the CALICE detectors in the CERN H6 test beam area, where data was taken in 2006 and 2007.

The AHCAL consists of 38 layers, each with a 1.6 cm thick steel absorber plate and a scintillator layer build out of
individual tiles housed in steel cassettes with a wall thickness of 2 mm, resulting in a total absorber thickness of 2 cm per layer. The lateral dimensions are roughly 1$\times$1 m$^2$, the total thickness amounts to 4.5 nuclear interaction lengths. The first 30 layers of the calorimeter have a high granular core of 10$\times$10 tiles with a tile size of 30$\times$30 mm$^2$, an outer core composed of 60x60 mm$^2$ tiles, and border tiles with a size of 120$\times$120 mm$^2$. The last 8 layers use only 60$\times$60 mm2 tiles in the core, and the large border tiles. In total, this amounts to 7608 channels. The light in each scintillator cell is collected by a wavelength shifting fiber, which is coupled to the SiPM. The SiPMs, produced by the MEPhI/PULSAR group \cite{Bondarenko:2000in}, have a photo-sensitive area of 1.1$\times$1.1 mm$^2$ containing 1156 pixels with a size of 32$\times$32 $\mu$m$^2$.

In this paper, a preliminary analysis of hadronic data, which explores the possibility to use the high granularity of the calorimeter to identify minimum-ionizing track segments within hadronic showers, is presented. This capability of imaging calorimeters can also be used for the cell-by-cell calibration of the detector in a collider experiment. First simulation studies within the ILD detector concept have been performed to further investigate this.

\section{Track Segments in Hadronic Showers}

\begin{figure}
\centering
\includegraphics[width=0.75\columnwidth]{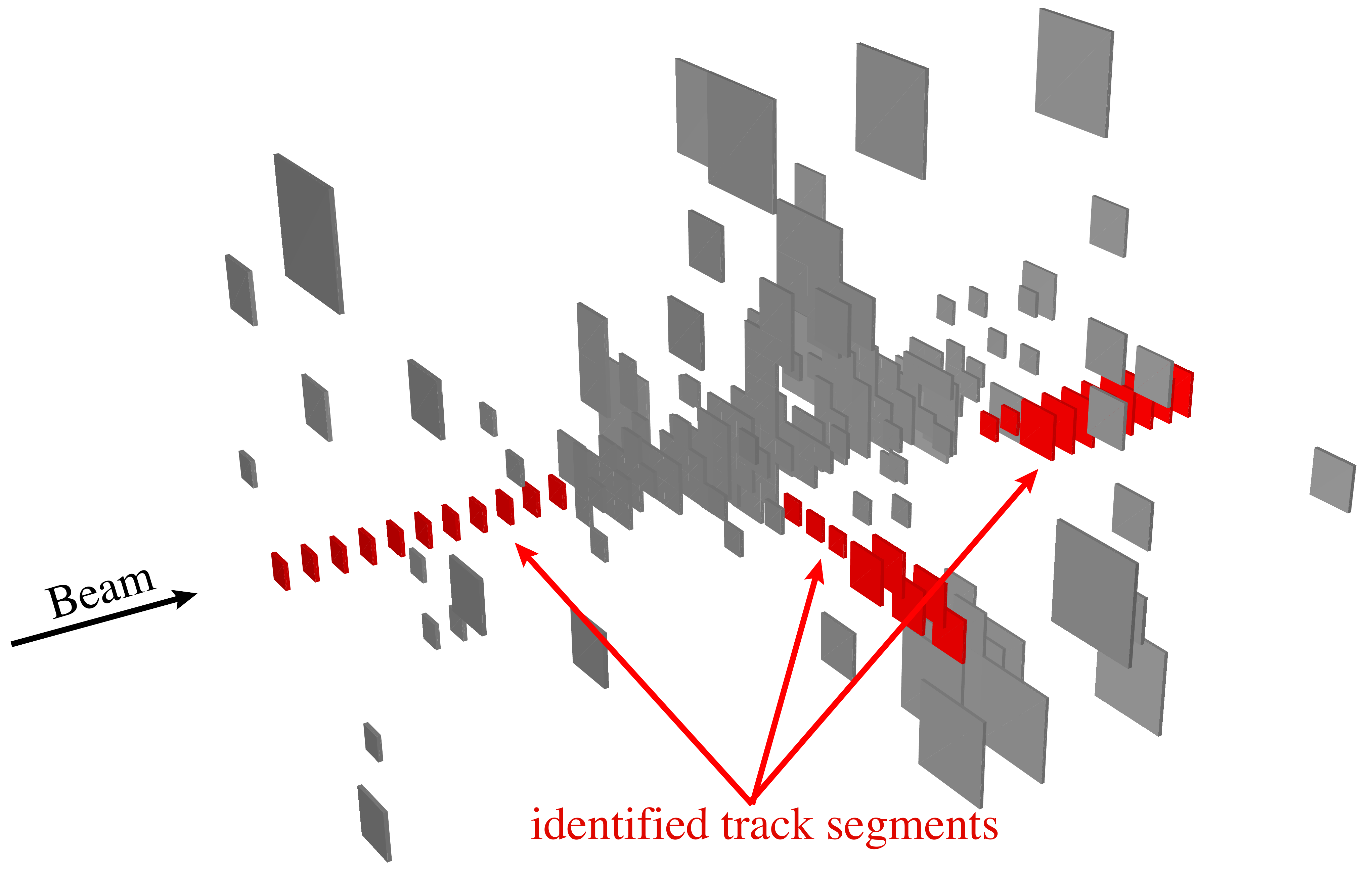}
\caption{Event display of a hadronic shower in the AHCAL initiated by a 25 GeV $\pi^-$. All cells in the calorimeter with an energy deposit above 0.4 MIP are shown, the identified track segments are highlighted in red. One of these tracks is the incoming $\pi^-$ before the first interaction. In addition two tracks of particles created in the shower are identified, one of which crosses over into the TCMT.}
\label{fig:EventDisplay}
\end{figure}

The high granularity of the active layers in the hadronic calorimeter and the cell-by-cell readout gives the CALICE detectors unprecedented imaging capabilities. This is exploited to study the topology of hadronic events in detail. Track segments created by charged particles produced within the hadronic shower can be identified, provided the particles travel an appreciable distance before interacting again and are separated from other activity in the detector.

\begin{figure}
\centering
\includegraphics[width=0.48\columnwidth]{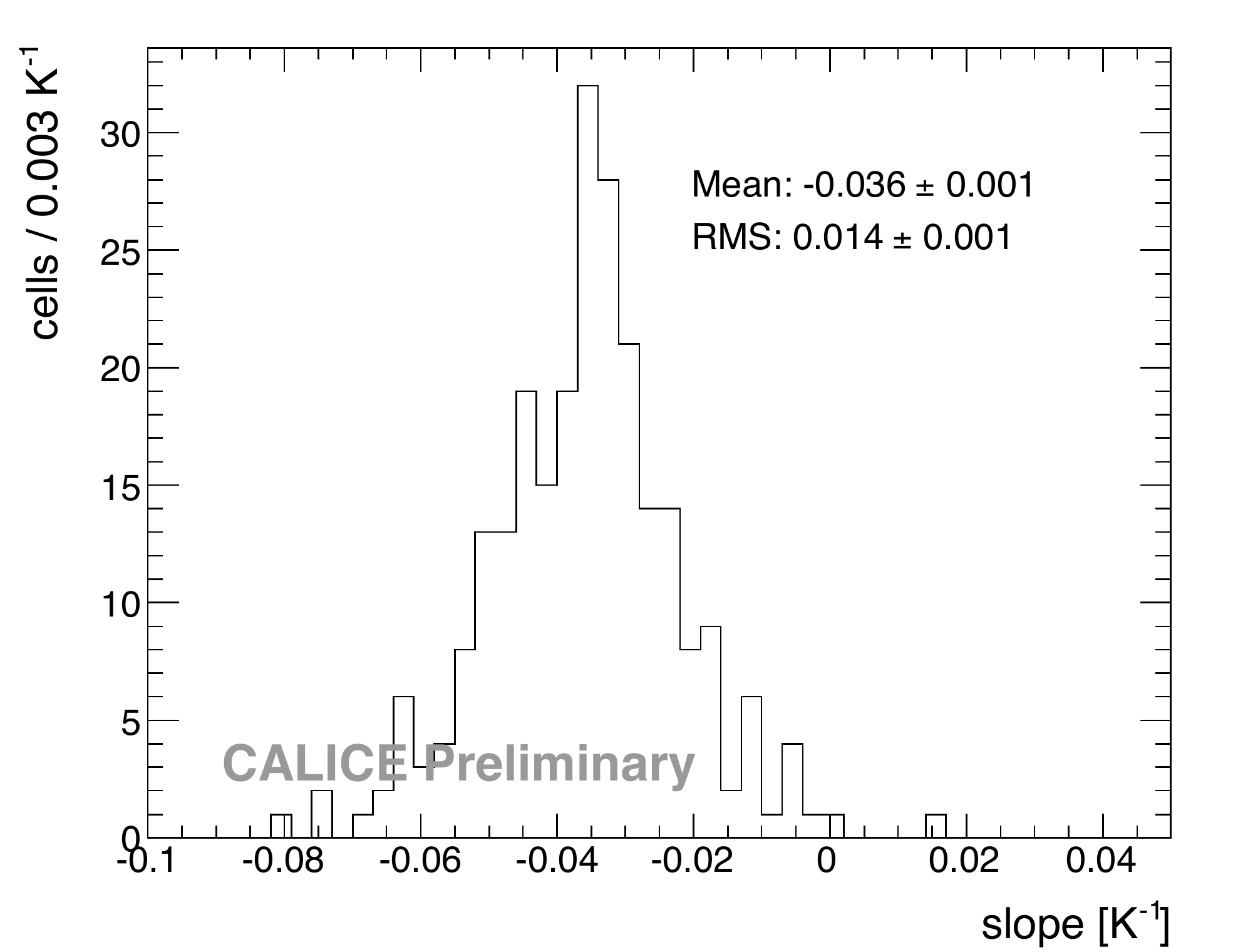}
\includegraphics[width=0.48\columnwidth]{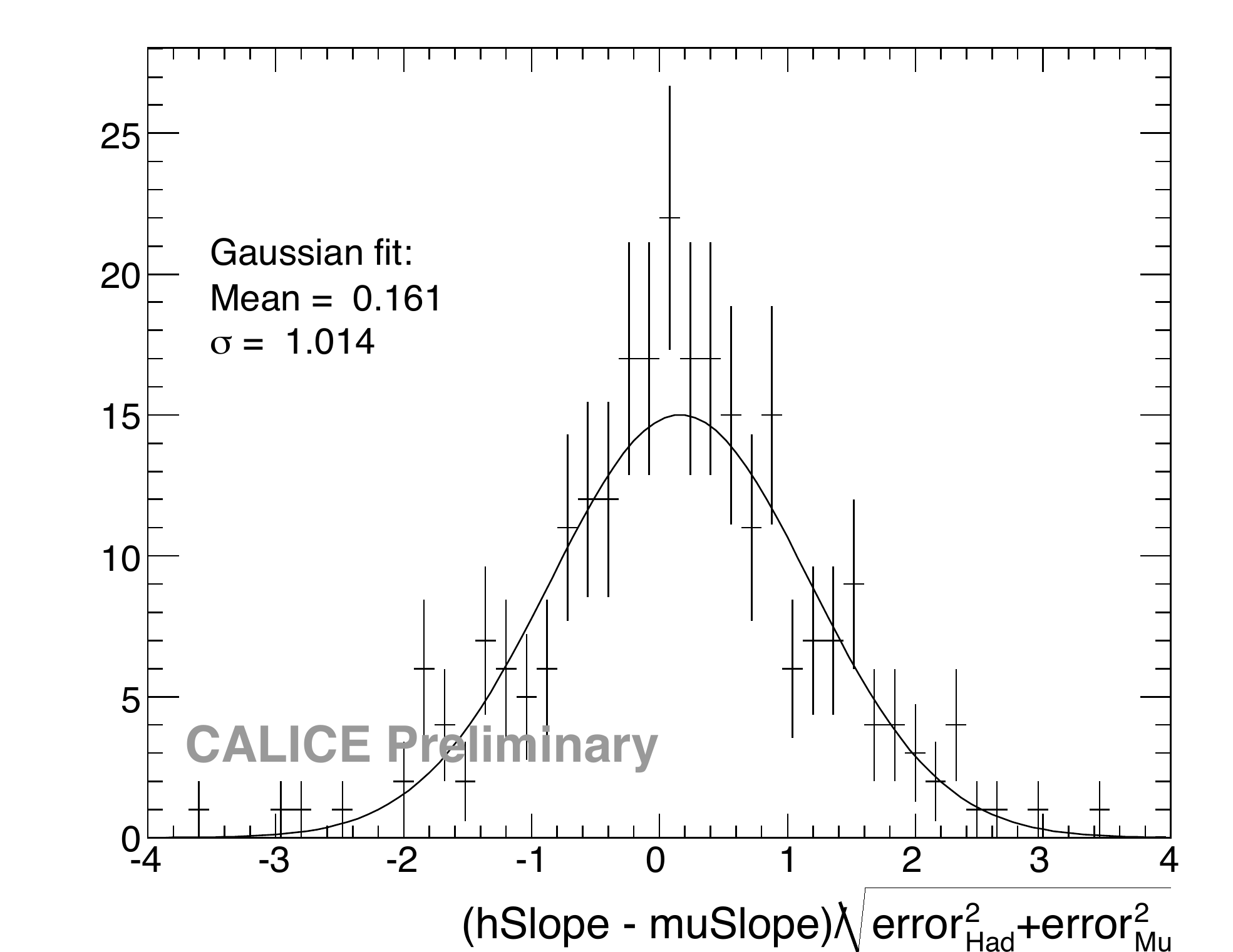}
\caption{{\it Left:} Distribution of temperature slopse of the cell response for track segments in hadronic showers. {\it Right:} Comparison of temperature slope determined from hadronic events and from muon events, demonstrating consistency of the two calibration methods.}
\label{fig:TempSlope}
\end{figure}

Figure \ref{fig:EventDisplay} shows an example event of a 25 GeV $\pi^-$ in the AHCAL, the other detectors of the CALICE setup are not shown. In this particular event the track of the incoming $\pi^-$ before the first interaction is seen, together with two additional track segments. Shown are all cells with a recorded energy above 0.4 times the energy corresponding to the most probable value of the energy loss of minimum ionizing particles. The response of all cells in the detector was calibrated as described in \cite{Simon:2008qj}. Track segments are found from isolated hits, cells with an energy deposit above a threshold of 0.4 MIP that do not have an energy deposit above threshold in the direct next neighboring cells in the same layer.

The track segments identified in hadronic showers have a high quality, and are suitable for detector calibrations via the extraction of the most probable value of the energy loss in each cell along the track. This feature was used to investigate the temperature dependence of the detector response, exploiting the large range of ambient temperatures in the presently available dataset. Further detail on the first preliminary studies with a subset of the available data can be found in \cite{Simon:2008qj}. Figure \ref{fig:TempSlope} shows the distribution of the temperature dependence of the cell response, determined from track segments in hadronic events. This distribution contains 250 detector cells, a small subset of $\sim$3\% of all channels, given by the available statistics, which are only significant in the center of the detector near the shower axis. Also shown is the comparison between the temperature dependence extracted with the novel method of track segments in hadronic events compared to the extraction from calibration data taken with muons. The distribution of the difference of the two methods, normalized by the measurement errors, shows good consistency of the two methods. This validates the method of extracting calibration quality data from hadronic events.

\section{Calibration Data from Hadronic Events: Prospects}

The cell-by-cell calibration of a complete hadronic calorimeter for the ILC, with approximately 5 million scintillator cells in the barrel calorimeter, is a formidable task. Cosmic muons will likely not be sufficient, since very high statistics are needed due to the high granularity, and the muon flux is limited in underground locations. In addition, the power-pulsing scheme used for all ILC detector electronics only allows signal sampling for less than 1\% of the time. Thus, alternative calibration mechanisms are necessary. To monitor short-term variations of the SiPM response, a LED calibration system is foreseen, which is already used in the CALICE AHCAL \cite{Simon:2008qj}. The study of long-term variations and the cell-by-cell intercalibration of the detector could potentially be performed using standard data events with the method of track segment identification, as discussed in the present paper. The quality of the track segments that can be identified within hadronic showers is sufficiently good that these can be used for a cell-by-cell calibration of the detector, based on the reconstructed value of the most probable energy deposit of these minimum-ionizing particles. 

\begin{wrapfigure}{r}{0.5\columnwidth}
\centerline{\includegraphics[width=0.49\columnwidth]{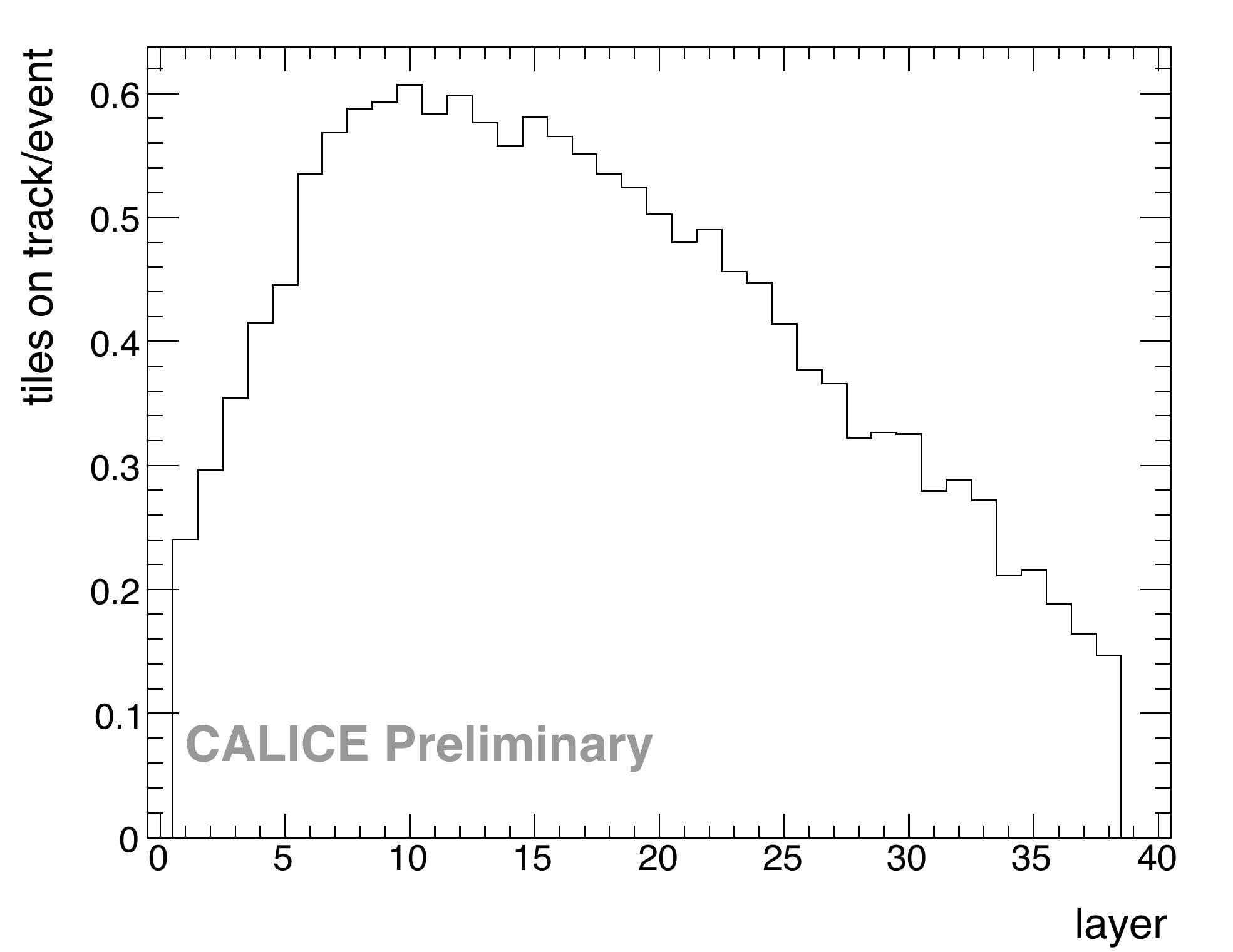}}
\caption{Number of cells crossed by identified hadronic track segments versus depth in the hadronic calorimeter.}
\label{fig:TrackDepth}
\end{wrapfigure}

A prerequisite for the use of hadronic events as a calibration tool are sufficient statistics also in the last layers of the calorimeters, where the expected particle multiplicity is significantly reduced due to the large amount of material in front. Figure \ref{fig:TrackDepth} shows the number of cells crossed by identified minimum-ionizing tracks per event as a function of the depth in the \mbox{AHCAL}. Taking a muon contamination of the beam of a few percent into account, this shows that still one in ten events with a 25 GeV hadron has a track in the last layer of the detector. The situation in the \mbox{CALICE} setup corresponds to a realistic collider detector, since there is a 30 $X_0$ silicon tungsten ECAL upstream of the AHCAL. The rise of the number of identified tracks in the first few layers of the detector is in part due to the tracking algorithm, which in this preliminary study can not track across detector boundaries and thus does not find tracks that cross from the ECAL into the HCAL. This affects the first six layers in particular, since a minimum track length of six layers is required in the algorithm. In addition, the higher particle density in the early parts of the shower makes tracking more difficult. Overall, on average 1.7 track segments with an average length of 10.5 layers are identified per event for 25 GeV $\pi^-$ \cite{Simon:2008qj}.

To evaluate the possibilities of tracking within the hadronic calorimeter of a future ILC detector, a first study on simulated data for the ILD detector concept has been performed. This detector concept uses a SiW ECAL and a scintillator tile HCAL, very similar to the CALICE setup discussed here. The tracking algorithm developed for the CALICE HCAL has been applied to simulated data, such as $e^+e^- \rightarrow q\bar{q}$ at $\sqrt{s}$ = 500 GeV and at the $Z^0$ resonance, to see if tracking is in principle possible in ILC data events. First results are encouraging, and yield similar track length distributions as observed in the CALICE data. Further optimization is necessary to account for the presence of a strong magnetic field, and for differences in the geometry. Currently, the total number of events needed to get sufficient statistics for a calibration of the first 20 layers of the calorimeter seems to be in the order of tens of millions when requiring at least $\sim$ 1000 identified tracks per cell. This excludes the use of data taken at $\sqrt{s}$ = 500 GeV due to the small $e^+e^-$ cross section at full ILC energy. At the $Z^0$ resonance, the cross section is much more favorable. However, the lower energy in combination with the high magnetic field reduces the statistics in the later layers of the detector in comparison to the situation in high energy events. The preliminary studies indicate the need of the order of 1 fb$^{-1}$ of $Z^0$ resonance data for a precise intercalibration of the first 15 layers of the HCAL barrel. The quality of the calibration in the later layers will decrease with decreasing statistics in those detector cells. By combining the data of several cells, the needed statistics can be reduced significantly, with the caveat of reduced granularity of the calibration. Such a strategy might be an attractive option to monitor possible long-term variations in the detector. 

These preliminary investigations suggest that track segments identified in the hadronic calorimeter in regular data events can indeed be used for the cell-by-cell calibration of the detector. However, the need for high statistics likely requires the availability of a large sample of $Z^0$ resonance data for this calibration approach. Possibilities for further improvements of the track identification, which would lead to a reductions of the required integrated luminosity for calibration using the hadronic track segments technique, are currently under study.

\section{Conclusion}

The CALICE collaboration studies highly granular calorimeters for detectors at the future International Linear Collider. We presented a first preliminary study of the tracking capability of the scintillator tile analog hadron calorimeter with SiPM readout. Due to the extremely high granularity, the identification and reconstruction of minimum-ionizing track segments of charged particles created within hadronic showers is possible. These track segments are used for detailed detector studies, such as the investigation of the temperature dependence of the response, yielding results consistent with those obtained with muon data. The track segments can also be used for a cell-by-cell calibration of the calorimeter at a future ILC detector. First proof-of-principle studies using standard data events simulated for the ILD detector concept suggest that track identification of sufficient quality is indeed possible. Tracking within hadronic showers could be a viable calibration strategy, however likely requiring a large $Z^0$ resonance dataset to obtain sufficient statistics in each detector cell.


\begin{footnotesize}

\bibliographystyle{h-physrev3.bst}

\bibliography{CALICE}

\end{footnotesize}


\end{document}